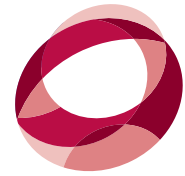

# Addressing the Unforeseen Harms of Technology - CCC Whitepaper
# June 2024

**Authors**[1]: Nadya Bliss (Arizona State University), Kevin Butler (University of Florida), David Danks (University of California, San Diego), Ufuk Topcu (University of Texas at Austin), and Matthew Turk (Toyota Technological Institute at Chicago)

**With support from:** Haley Griffin (CCC)

**How to cite this report:** Bliss, N., Butler, K., Danks, D., Topcu, U., & Turk, M. (2024). *Addressing the Unforeseen Harms of Technology*. Computing Community Consortium. http://cra.org/ccc/addressing_unforeseen_harms_of_technology_ccc_whitepaper

---

[1]The CCC Council formed a task force on Addressing the Unforeseen Deleterious Impacts of Technology (AUDIT) to investigate possible harmful consequences of computing technology, to what extent these outcomes could have been mitigated or avoided, and who should be responsible for negative impacts. The group also spoke with multiple external experts to augment their research experience on the topics.

## Introduction

Recent years have seen increased awareness of the potential significant impacts of computing technologies, both positive and negative. This whitepaper explores how to address possible harmful consequences of computing technologies that might be difficult to anticipate, and thereby mitigate or address. It starts from the assumption that very few harms due to technology are intentional or deliberate; rather, the vast majority result from failure to recognize and respond to them prior to deployment. Nonetheless, there are concrete steps that can be taken to address the difficult problem of anticipating and responding to potential harms from new technologies.

## Unforeseen vs. Unforeseeable

Advances in computing technologies have brought enormous benefits to people's lives, but they have also led to significant individual and societal harms. As these technologies become increasingly ubiquitous and powerful, we should expect the potential benefits and harms to grow as well. These shifts raise crucial questions about the foreseeability of impacts from the work of computing researchers and developers, as it is much easier to promote benefits and mitigate harms when they can be anticipated. We can ensure wide access (if beneficial), establish guardrails (if problematic), and much more, but only if we actually foresee how the computing technology will be designed, developed, and deployed in the real world [NASEM, 2022].

In some cases, it is easy to anticipate the impacts of a new technology. For example, the first-order impacts of a faster processor can usually be modeled and estimated. At the same time, more complicated impacts can be much harder to anticipate; for instance, we might encounter a *Jevons paradox* in which increased efficiency leads to increased utilization, thereby undoing the positive benefits of the efficiency gains. As a practical example, autonomous vehicles are likely to be more efficient per vehicle mile traveled, but if they lead to an increase in total vehicle miles traveled, then emissions could actually rise when autonomous vehicles are introduced [Kalra & Groves, 2017; Geary & Danks, 2019]. These complexities in anticipating benefits and harms only grow as the capabilities and sophistication of our computing technologies increase. And of course, matters are exponentially more complicated when we consider *research* on computing technologies, as the intellectual and temporal gaps between research and implementation can be vast [NASEM, 2020].

> *…we have a societal and ethical obligation to anticipate and address the foreseeable impacts of our efforts to bring new technologies into the world.*

Despite these challenges, we have a societal and ethical obligation to anticipate and address the foreseeable impacts of our efforts to bring new technologies into the world. Companies and organizations typically explore ways to consistently maximize the benefits of technologies that they produce, but they do not have the same record of anticipating the potentially harmful impacts of their new computing technologies. Consider a few examples: Mortgage approval systems were deployed with an understanding of how they could increase profit for lenders, but not how they could increase inequality in access to financial resources. Many people failed to anticipate the ways that social media would change social interactions for the worse. Automated hiring systems have unintentionally codified sexist and racist practices. There are many more cases of unforeseen harm and challenges.

We might hope that failures to anticipate harm occur only because of the complexity of ways in which technologies can interact with and shape communities and societies. However, there are often incentives and institutional structures that create reasons to avoid anticipating. That is, many problematic effects are arguably "willfully unforeseen," rather than justifiably unforeseen. Appropriate and thorough analyses would have identified those potential harms, but such analyses were not done. In such cases, we cannot simply point to our personal or organizational failures to anticipate harm to absolve ourselves from blame. We are responsible for the impacts that we *should* have foreseen, even if we did not actually foresee them in this particular situation. And so we need to recognize and address the barriers to actually understanding the impacts of the computing technologies that we create.

On the incentives side in industry, many companies and organizations reward people for "writing code" or other activities on the basis of solely "local" benefits, rather than more holistic assessments of all impacts. That is the incentives for an individual employee or team all point towards a focus on potential *benefits* to the exclusion of other potential impacts. Meanwhile, in academia, tenure and promotion depend on publications and grants, where there is little incentive to emphasize potential harms or problems. The temptation to focus on benefits is also heightened by the typical distance between academic research and technology deployment. In all of these cases, it is little surprise that people do not spend much time thinking about what could go wrong. The harms often are not foreseen, even though they were foresee*able*.

> *...many problematic effects are arguably "willfully unforeseen," rather than justifiably unforeseen.*

On the institutional side, whether in industry or academia, computing technologies–both research and development–are often created by people who are far removed from key stakeholders. Many harms from new computing technologies are easily seen by the impacted communities, but not necessarily by those tasked with creating or researching that technology [NASEM, 2022; Gebru et al., 2024]. However, direct engagement with impacted communities, whether through minimal interactions such as focus groups or richer interactions such as co-design, is not systematically part of all projects to create new computing technology. We need to be talking with those who will interact directly with the technology, but those connections can be rare-to-nonexistent in many situations.

One might despair at this point, as the challenge of anticipating the benefits and harms of computing technologies appears too difficult, whether technically or institutionally. Although we face a difficult task, there are various methods and organizational designs that are being developed and tested to help us all do a better job of understanding likely impacts [NASEM, 2022]. These approaches range from practices that identify possible harms (e.g., red-teaming), to changes in organizational cultures (e.g., naming Chief (Responsible) AI Officers to lead these efforts, or encouraging academics to engage with potentially impacted communities), to different policy or regulatory approaches (e.g., holding companies liable for certain harms). Of course, even the best-intentioned efforts might fall short, and so we must also consider ways to address harms regardless of whether or not they were foreseeable.

## Responding through Design

Mitigating risks in the design phase is a common tenet of software engineering. A 2002 report from the National Institute of Standards and Technology found that the cost of correcting an error once a product is released is 30X higher than making the correction if it is found in the design phase [Tassey, 2002]. It is cheaper to fix a problem earlier in the development cycle. And some problems cannot be fixed if they are noticed too late in the development process; some harms cannot be undone or mitigated. These lessons apply not only to errors; they apply to properties such as the security of a system as well. As early as 1972, James P. Anderson wrote in an Air Force Technical Report that "merely saying a system is secure will not alter the fact that unless the security for a system is designed in at its inception, there are no simple measures to later make it secure" [Anderson, 1972].

Unfortunately, while these tenets have been known for decades, they have often not been practiced. Mark Zuckerberg famously declared in 2012 that Facebook had a saying: "Move fast and break things" [Zuckerberg, 2012]. Security has often been treated in a similarly cavalier fashion, with the design paradigm of "penetrate-and-patch," where software vulnerabilities are fixed after a compromise is found through a patch to the software system, representing an expected mode of operation [McGraw, 1998]. For example, Microsoft has maintained "Patch Tuesday" updates to its core software for years [Neumann, 2006] to address security updates; this model continues on modern devices, from computers and laptops to smartphones and IoT devices.

> *…do we want to address harmful, complex issues arising from large AI systems only after they have occurred, or should we put in the time and effort to ensure that secure and private designs are in place to mitigate future concerns?*

How can we ensure that systems are responsible by design? The answer can be challenging, as designers need to take the time to discuss and anticipate negative consequences. Even if these consequences are addressed, there may still be future cases, uses, or contexts that need to be accounted for in the design. For example, the MS-DOS operating system was designed based on the system requirements of the IBM PC in the early 1980s, when computers were designed to be single-user machines used in isolation. When computers became multi-user and, more importantly, connected to networks such as the Internet, the usage model changed, which meant that the risks such a computer would face also changed in ways that made MS-DOS unsuitable.

Therefore, designers must be as forward-thinking as possible so that potential consequences are not unforeseen. There are many lessons that we can draw from the past to guide design and ensure that the unforeseen can still be addressed. Let us consider privacy as an example. By changing the system to require the user to explicitly opt-in to information collection, and setting the default to opt-out, information exposure is reduced, which can prevent information loss if a central data store is breached, even if the method of compromise is currently unforeseen. Such a design may conflict with other stakeholders in a solution, such as a business unit that seeks to monetize data collected from a system. However, being aware of the risks and metrics, as well as externalities engendered by risks, can allow for informed discussion that incentivizes security by design before a system is deployed [Bliss et al., 2020].

We are facing significant questions about risks with the development and deployment of AI systems. For example, generative AI has the potential to enable entirely new fields and fundamentally change the face of industry, but these systems also change the threat landscape, providing both attackers and defenders with new capabilities [Barrett et al., 2023]. While there has been some discussion of installing guardrails around AI systems while we seek to understand the threat surface they represent, others have proposed charging forward, not wanting to give up the strategic value of being first past the post with technological advances that could potentially change industry or even society as a whole. We must ask ourselves: do we want to address harmful, complex issues arising from large AI systems only after they have occurred, or should we put in the time and effort to ensure that secure and private designs are in place to mitigate future concerns, especially given the potential for catastrophic risk from these systems [Hendrycks et al., 2023]? The calculus is not always easy in this situation given the effort involved in designing a system for safety from scratch. The key difference, of course, is that a system that merely fixes previous vulnerabilities will still be susceptible to vulnerabilities in the future, while designing for safety can make a system resilient to future threats.

The choices that we make now can have implications for every aspect of society in the future.

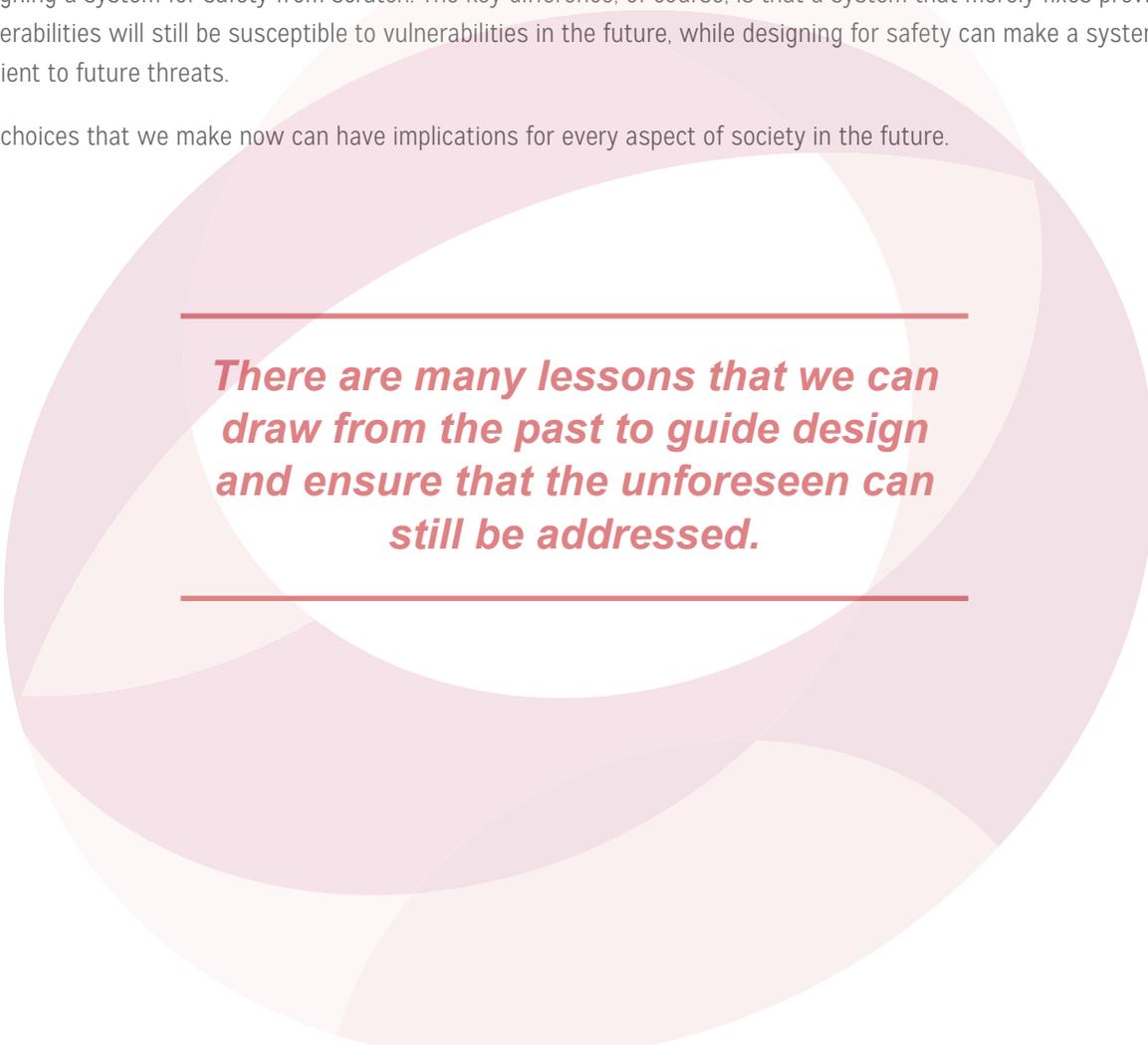

*There are many lessons that we can draw from the past to guide design and ensure that the unforeseen can still be addressed.*

# Notes

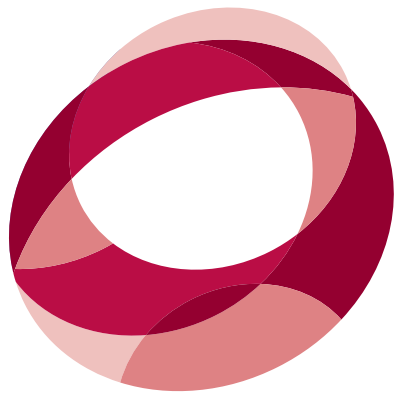

Computing Research Association
1828 L St., NW, Ste. 800
Washington DC, 20036